\documentclass{PoS}

\usepackage{lineno}
\title{IceCube Search for Galactic Neutrino Sources based on 
HAWC Observations of the Galactic Plane}

\ShortTitle{Galactic Neutrino Sources \& HAWC}

\author{
The IceCube and HAWC Collaborations\footnote{For collaboration list, see PoS(ICRC2019) 1177.}\\
{\itshape \href{http://icecube.wisc.edu/collaboration/authors/icrc19_icecube}{http://icecube.wisc.edu/collaboration/authors/icrc19\_icecube}, \href{https://www.hawc-observatory.org/collaboration/icrc2019.php}{https://www.hawc-observatory.org/collaboration/icrc2019}}\\
E-mail: \email{akheirandish@icecube.wisc.edu, jwood@icecube.wisc.edu}
}

\abstract{We present a search in IceCube data for neutrino emission from Galactic TeV gamma-ray sources detected by the HAWC gamma-ray observatory. HAWC serves as the excellent instrument to complement IceCube with its energy range extending to very high energies. Assuming that the highest energy photons originate from the decay of pions, rather than from accelerated leptons, the very high energy gamma-rays observed by HAWC are expected to be correlated with neutrinos. Using eight years of IceCube data, we report on two analyses that investigate a possible neutrino--gamma ray correlation. The first is a stacked analysis of identified HAWC point sources and the second is a template method which accounts for the full morphology of HAWC sources, including their measured extension.\\

\vspace{4mm}
{\bfseries Corresponding authors:}
\speaker{Ali Kheirandish}$^{1}$, Joshua Wood$^{1}$\\
{$^{1}$ \itshape Department of Physics \& Wisconsin IceCube Particle Astrophysics Center, University of Wisconsin, Madison, WI 53706, USA}\\
}

\FullConference{36th International Cosmic Ray Conference -ICRC2019-\\
		July 24th - August 1st, 2019\\
		Madison, WI, U.S.A.}
		
\begin{document}
\section{Introduction}\label{sec:intro}
In the search for the origin of very high energy particles, the Galactic plane provides the richest environment to study the nature and mechanism of particle acceleration. Galactic cosmic ray accelerators are believed to produce cosmic rays (CRs) up to several PeV, the "knee" in the CR spectrum. The CRs interact with the dense environment in the Milky Way, and produce charged and neutral pions. The pions subsequently decay leading to high-energy neutrinos and gamma-rays which retain around 5\% and 10\% of the energy, respectively, from the primary cosmic ray \cite{PhysRevD.76.123003}. The locations near such accelerators are therefore excellent candidates to search for both very high energy gamma-rays and neutrinos with energies extending to several hundred TeV. 

Simultaneous production of neutrinos and gamma rays opens up the opportunity for multimessenger searches to find the origin of very high energy CRs. Multimessenger searches are essential for identifying these sources since the detection of a source in gamma-rays alone cannot reveal whether hadrons are being accelerated to very high energy. This is because very high energy gamma-rays can also originate from the acceleration of electrons. In contrast, high energy neutrinos can only be generated where hadrons are present. Therefore, a high-energy neutrino observation of a source will provide a smoking gun for the identification of a Galactic CR accelerator.

The very high energy gamma-ray survey of Galactic sources by the Milagro Collaboration \cite{Abdo:2006fq} revealed the brightest gamma-ray sources in the Northern sky. Promising sources were identified in the Milagro sky map based on their spectra and early estimations implied their likely observation within the initial operation years of IceCube \cite{Halzen:2008zj,GonzalezGarcia:2009jc}. Additional observations by Milagro and Imaging Air Cherenkov Telescopes (IACTs) provided more insight on the high-energy gamma-ray emission from the Milky Way. However, the prospect for the observation of these sources in IceCube became less clear with the discrepancy in the gamma-ray flux and the extension of these sources reported by different observatories \cite{Gonzalez-Garcia:2013iha, Halzen:2016seh}. This necessitated a deeper survey of the gamma-ray sky above 10 TeV which is now being provided by the High Altitude Water Cherenkov (HAWC) Observatory \cite{Abeysekara:2017hyn}.
\section{IceCube \& HAWC Synergy}\label{sec:ic-hawc}
The IceCube Neutrino Observatory \cite{Aartsen:2016nxy} is a high-energy neutrino detector located within the Antarctic ice sheet at the Amundsen-Scott South Pole Station. The detector consists of an array of 86 vertical strings instrumenting 1 km$^3$ of ice with optical sensors that record Cherenkov light emitted by high-energy charged particles passing through the ice. High-energy muon neutrinos interact with the ice to produce relativistic muons that may travel many kilometers, creating a track-like series of Cherenkov photons recorded when they pass through the array. These events range in energy from 0.2 TeV to 1 EeV and allow for the reconstruction of the original neutrino direction with a median angular uncertainty of 0.5$^\circ$ for a neutrino energy of around 30 TeV. The field-of-view of IceCube spans the entire sky because neutrinos at these energies can pass through the Earth. However, its best sensitivity for resolving sources lies in the Northern hemisphere where charged particle backgrounds, which are created locally by CRs interacting in the atmosphere, are completely removed by absorption in the Earth.

The HAWC Observatory is a TeV gamma-ray observatory that is ideally suited to measure sources that may also be identifiable by IceCube for several reasons. First, its location at 19$^\circ$ North and wide field-of-view provide a survey of the Northern sky where backgrounds in IceCube are lowest. Second, the broad energy coverage of the HAWC observatory from 0.3 TeV to 100 TeV overlaps with that of IceCube. Lastly, its angular resolution is better than that of IceCube at 0.2 deg for photon energies near 10 TeV.
\section{HAWC Observations of the Galactic Plane}\label{sec:hwc}
The first observations of the Galactic plane using the fully realized HAWC detector were published in the 2HWC catalog \cite{Abeysekara:2017hyn}. This catalog identified 39 gamma-ray sources at TeV energies in 507 days of data, 19 of which were more than 0.5$^\circ$ away from any previously detected TeV source. While some of these sources are firmly identified as pulsar wind nebulae (PWN) where the TeV gamma-ray emission originates from a population of high energy electrons, most have no firm identification. Additionally, in some cases the identified PWN counterpart can only explain a portion of the flux measured by HAWC \cite{Alfaro:2018xep}.

Sources identified by HAWC without clear PWN counterparts or where the counterpart only explains a portion of the flux measured by HAWC may represent Galactic accelerators with hadronic production of gamma-rays. Therefore, our work focuses on the set of HAWC sources from the 2HWC catalog with emission that cannot be explained entirely by a pulsar wind nebula counterpart. We note though that others have proposed the concept of TeV halos, which may provide a leptonic emission mechanism for the majority of sources seen by HAWC even in cases where a pulsar counterpart cannot be identified \cite{Linden:2017blp}.

\begin{figure}[ht!]
\centering
\includegraphics[width=6in]{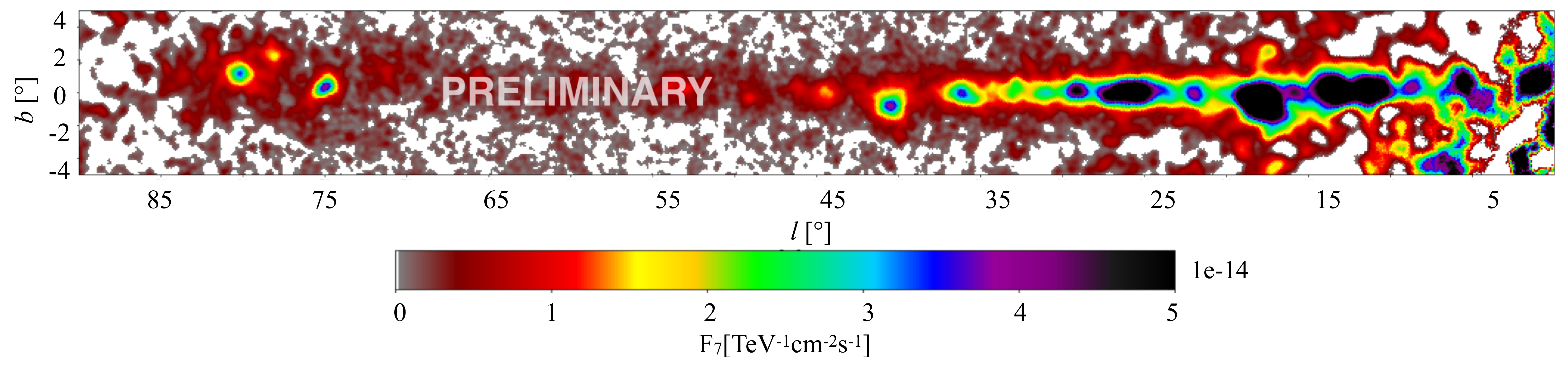}
\caption{Map of the flux at 7 TeV, $F_7$, measured in the Galactic plane by HAWC for a point source hypothesis at each location using fixed E$^{- 2.7}$ spectral weights.}
\label{fig:hawc_flux_map}
\end{figure}

After the 2HWC catalog was published, HAWC has produced an updated sky map using the same analysis techniques from the catalog with 1128 days of data (Figure \ref{fig:hawc_flux_map}). We use this map as the corresponding data set within our study.
\section{Searching for Neutrino Emission}\label{sec:ic} 
We search for neutrino emission associated with the very high energy gamma-rays observed by HAWC. If high-energy gamma-rays detected by HAWC are produced from decay of pions, high-energy neutrinos should inevitably accompany them. Given the synergy of HAWC and IceCube, as discussed in Section \ref{sec:ic-hawc}, a detection of neutrinos in association with HAWC sources would provide evidence for the acceleration of cosmic rays from these Galactic sources. In this study, we use eight years of muon neutrino track events from the Northern sky, a data set that was previously used in a search for neutrino point sources \cite{Aartsen:2018ywr}.

\begin{figure}[ht!]
\centering
\includegraphics[width=6in]{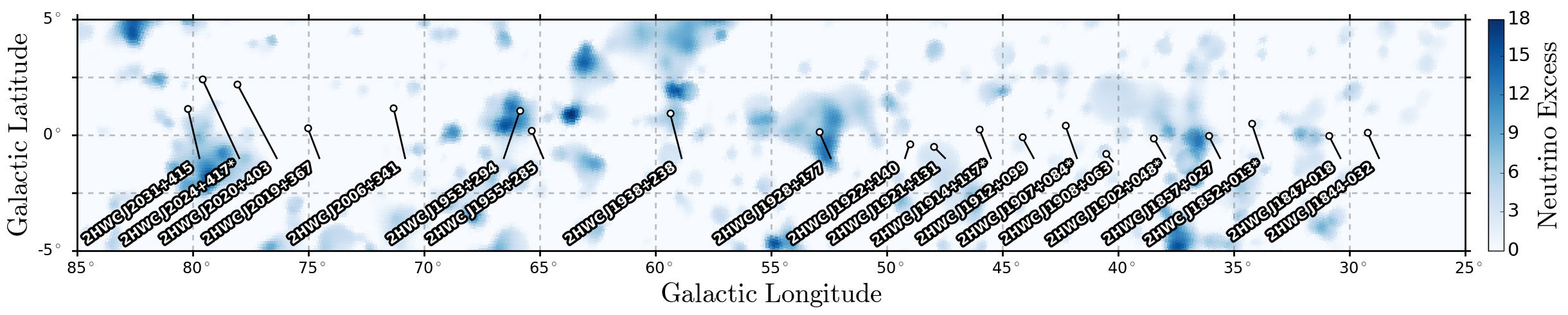}
\caption{Map of the neutrino excess in IceCube with HAWC sources overlaid.}
\label{fig:stacked_overlay}
\end{figure}

We incorporate a stacking likelihood method (see \cite{Achterberg:2006ik} for details) and test five different hypotheses for the correlation of gamma-rays and neutrinos. 
The first search is a stacking search for neutrino emission from HAWC sources that are not identified as PWN \footnote{We have excluded PWN in this search for two reason: first, high-energy emission from PWN is generally understood to be leptonic. Second, a dedicated search by the IceCube collaboration has examined neutrino emission from TeV PWN \cite{qliu2019}.}. This focuses on 20 sources shown in Fig. \ref{fig:stacked_overlay}. 

\begin{figure}[ht!]
\centering
\includegraphics[width=6in]{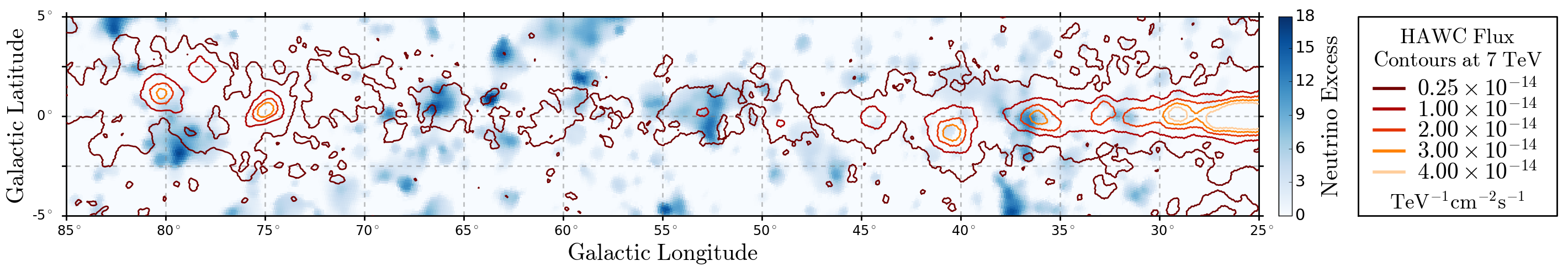}
\caption{Map of the neutrino excess in IceCube with HAWC flux contours overlaid.}
\label{fig:flux_contour_overlay}
\end{figure}

For the four other searches, we use the morphology of the gamma-ray emission as reported by the HAWC Collaboration. For this purpose, we incorporate the high-energy gamma-ray flux morphology as shown in Fig. \ref{fig:flux_contour_overlay} and use the gamma-ray emission at each point to weight the stacking likelihood. We do this search by considering the whole plane seen by the Northern sky muon neutrino sample (Dec. $> -5^\circ$) as well as for three regions defined {\em a priori}{: Cygnus region, and the areas surrounding 2HWC J1908+06, and 2HWC J1857+027. These are star forming regions with high levels of gamma-ray activity and young stars, and have been historically identified as potential neutrino emitters \cite{Beacom:2007yu,Halzen:2016seh}}.

The results of these likelihood tests are summarized in Table \ref{tab:results}. The observed excess of signal neutrinos found in these tests is not statistically significant for claiming evidence of neutrino emission associated with the gamma-ray emission measured by HAWC. The most significant test corresponds to the region surrounding 2HWC J1857+027 and finds an excess of 36.7 neutrinos with a pre-trial p-value of 2\%. 

\begin{figure}[t!]
\centering
\includegraphics[width=0.6\linewidth]{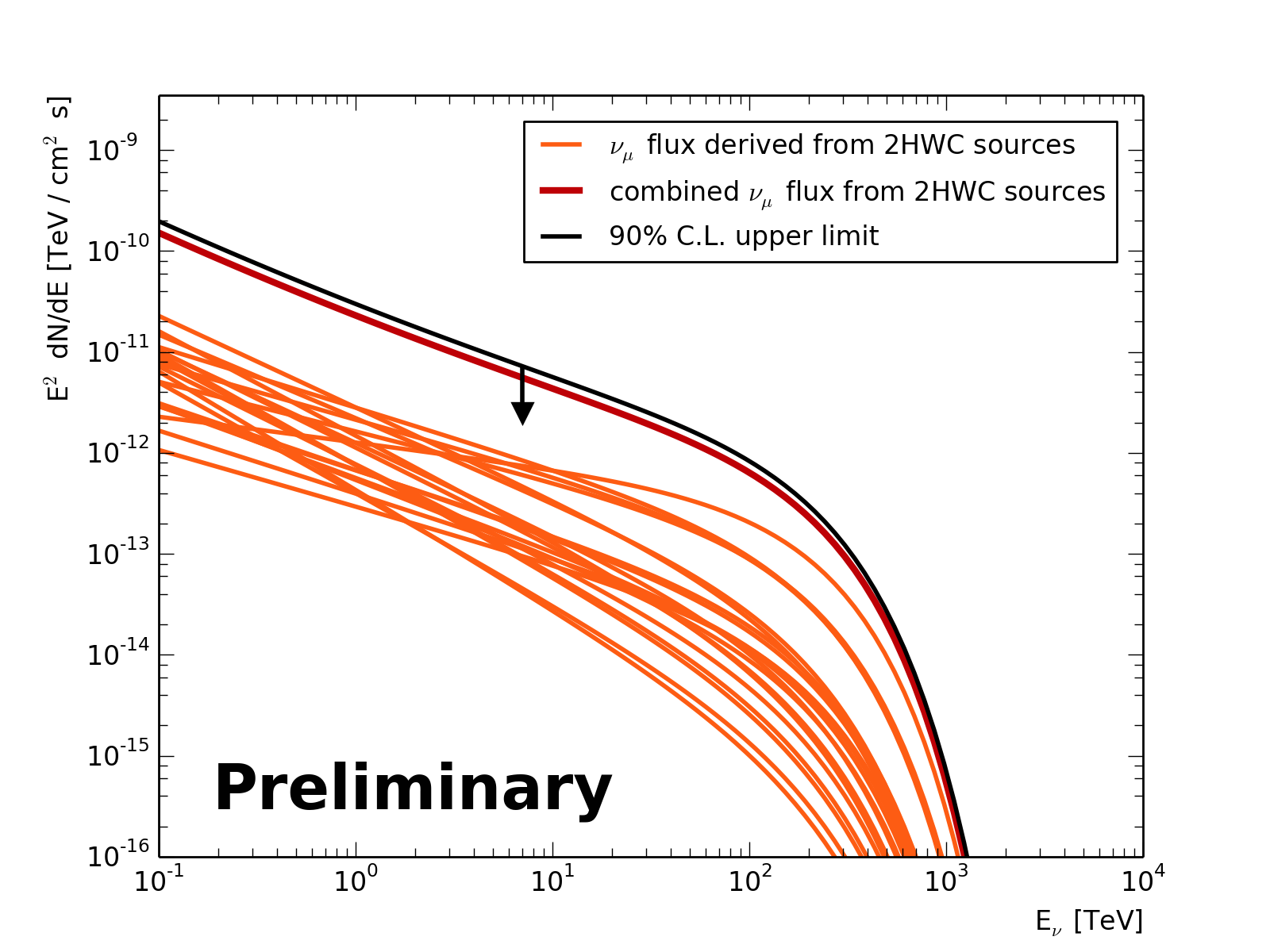}
\caption{Upper limit (90\% C.L.) on the flux of high-energy muon neutrinos (black) for the stacking search of non-PWN sources in the 2HWC catalog. The projected muon neutrino fluxes (thin orange) represent the expected flux from each source assuming that the high-energy gamma ray flux measured by HAWC is from hadronuclear interactions following \cite{Kappes:2006fg}. The combined flux (red) shows sum of the individual fluxes.}
\label{fig:ul_stacked_plane}
\end{figure}

\begin{figure}[ht!]
\centering
\includegraphics[width=0.49\linewidth]{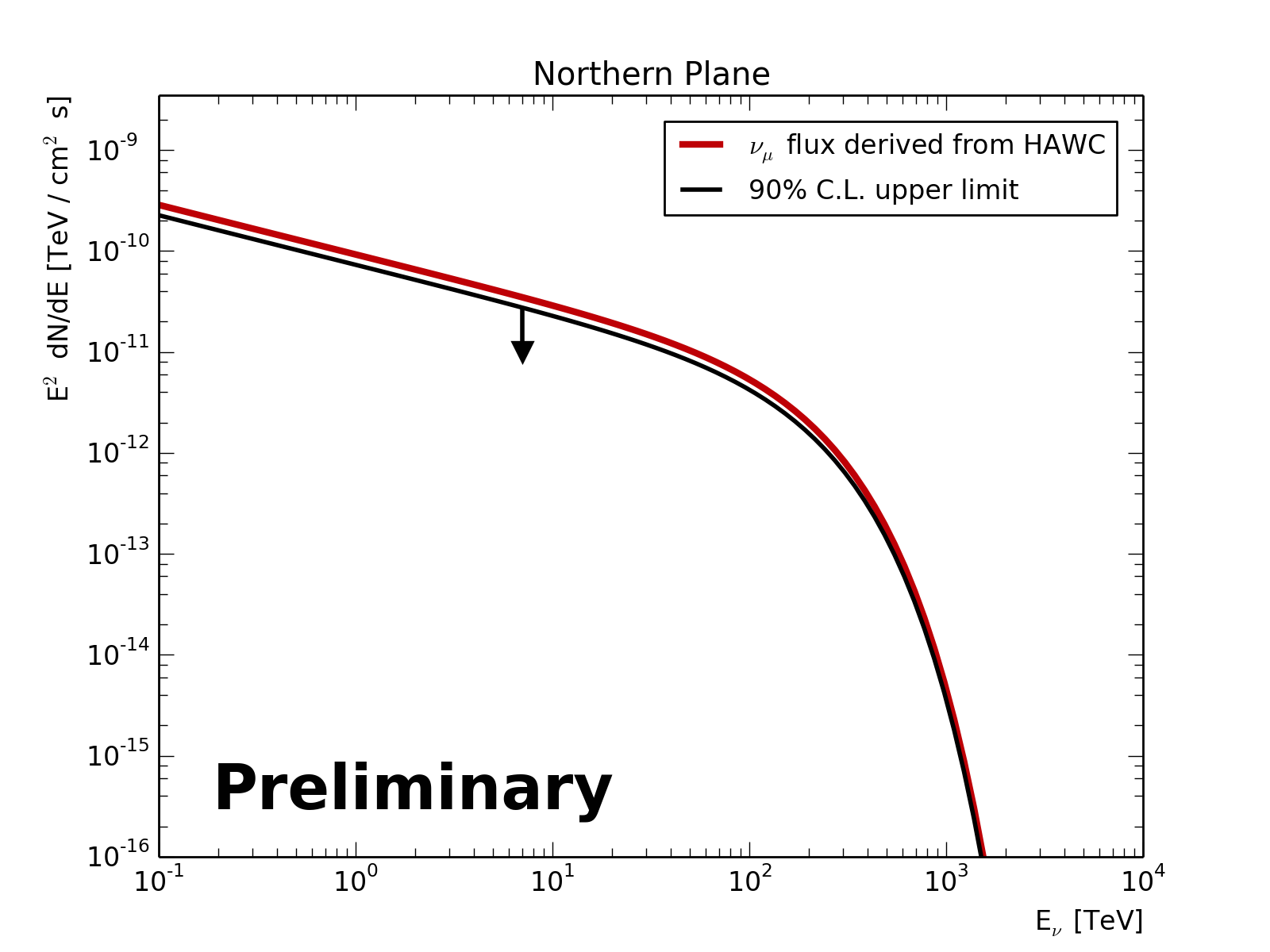}
\includegraphics[width=0.49\linewidth]{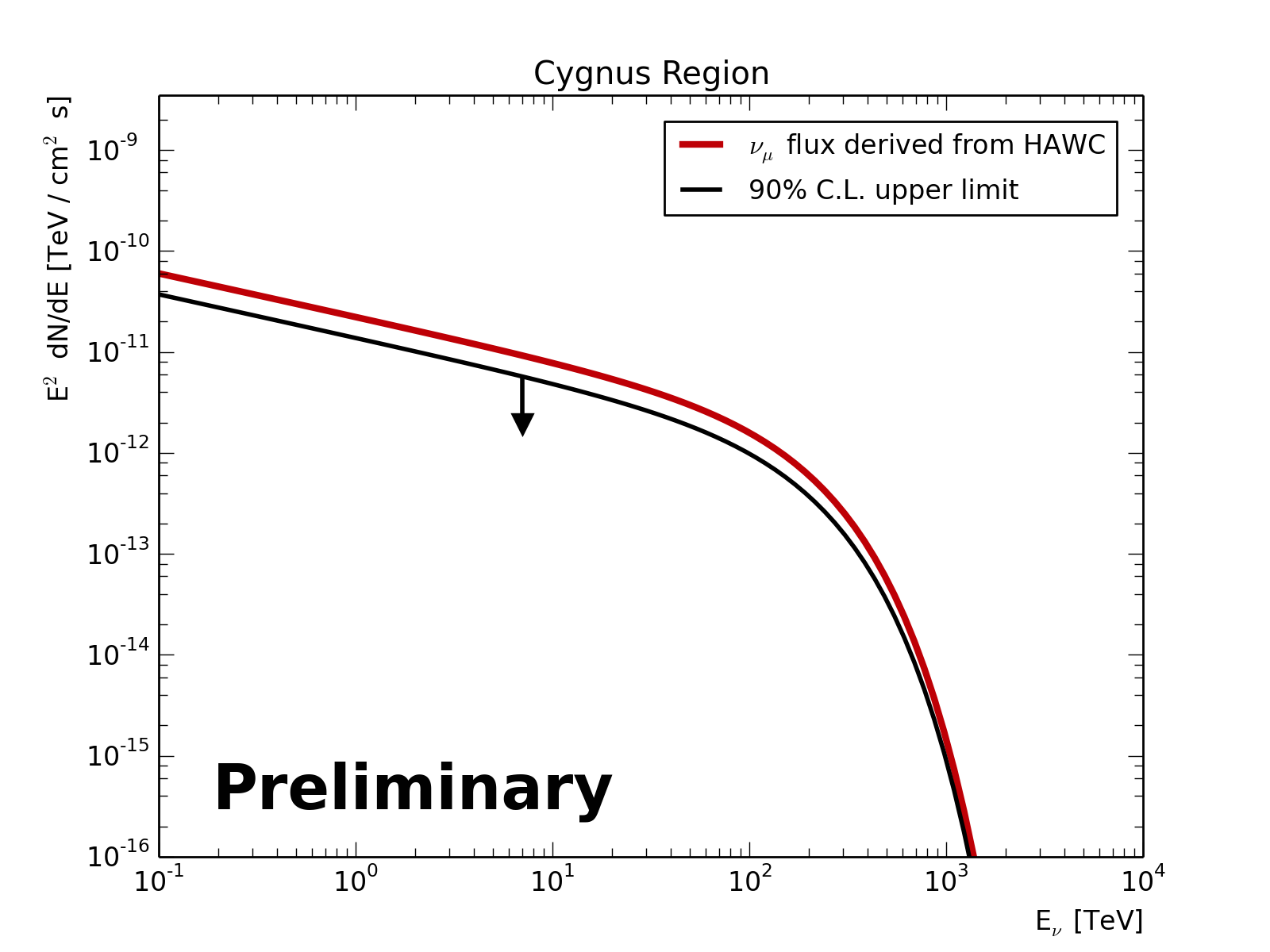}
\includegraphics[width=0.49\linewidth]{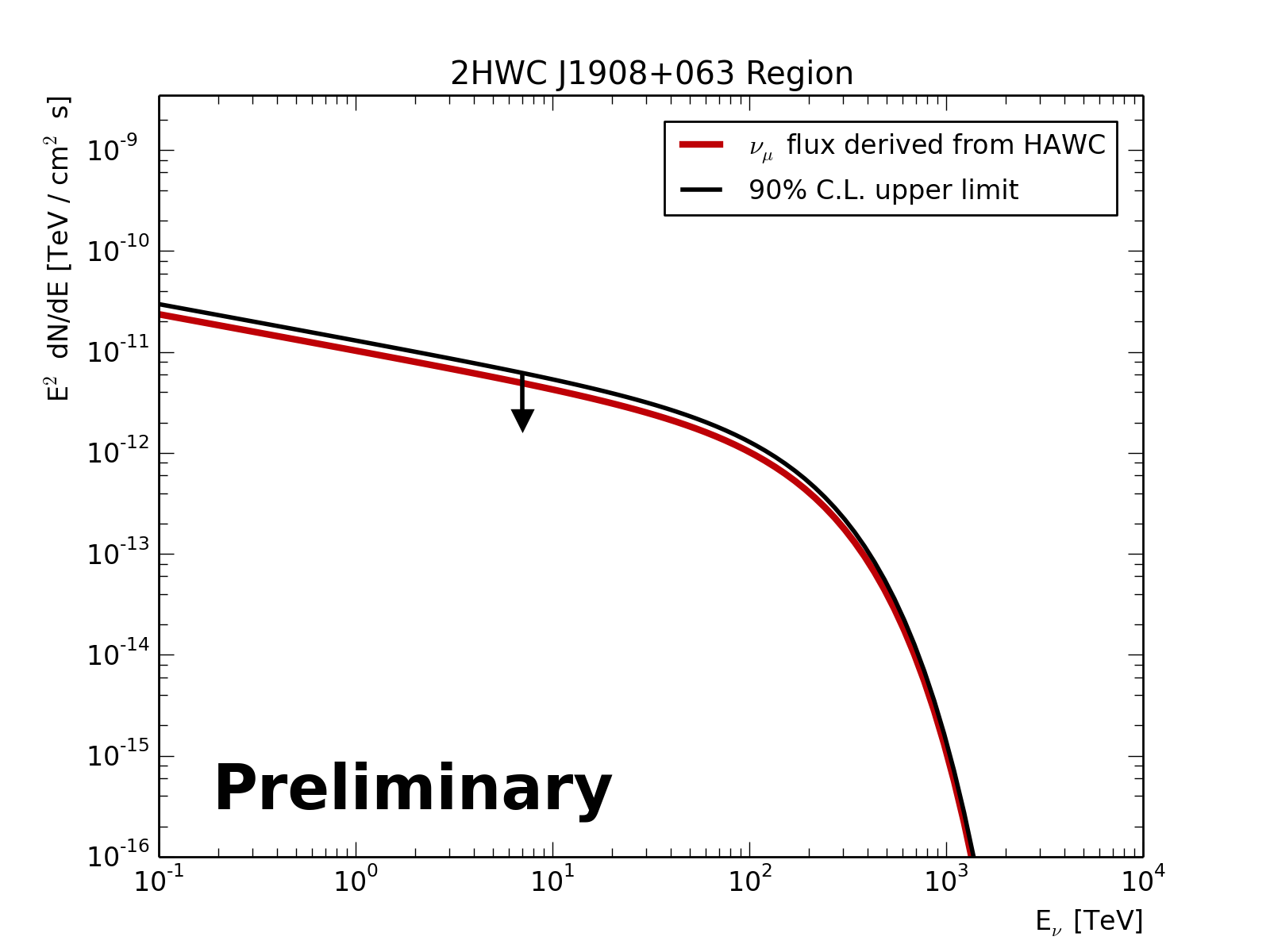}
\includegraphics[width=0.49\linewidth]{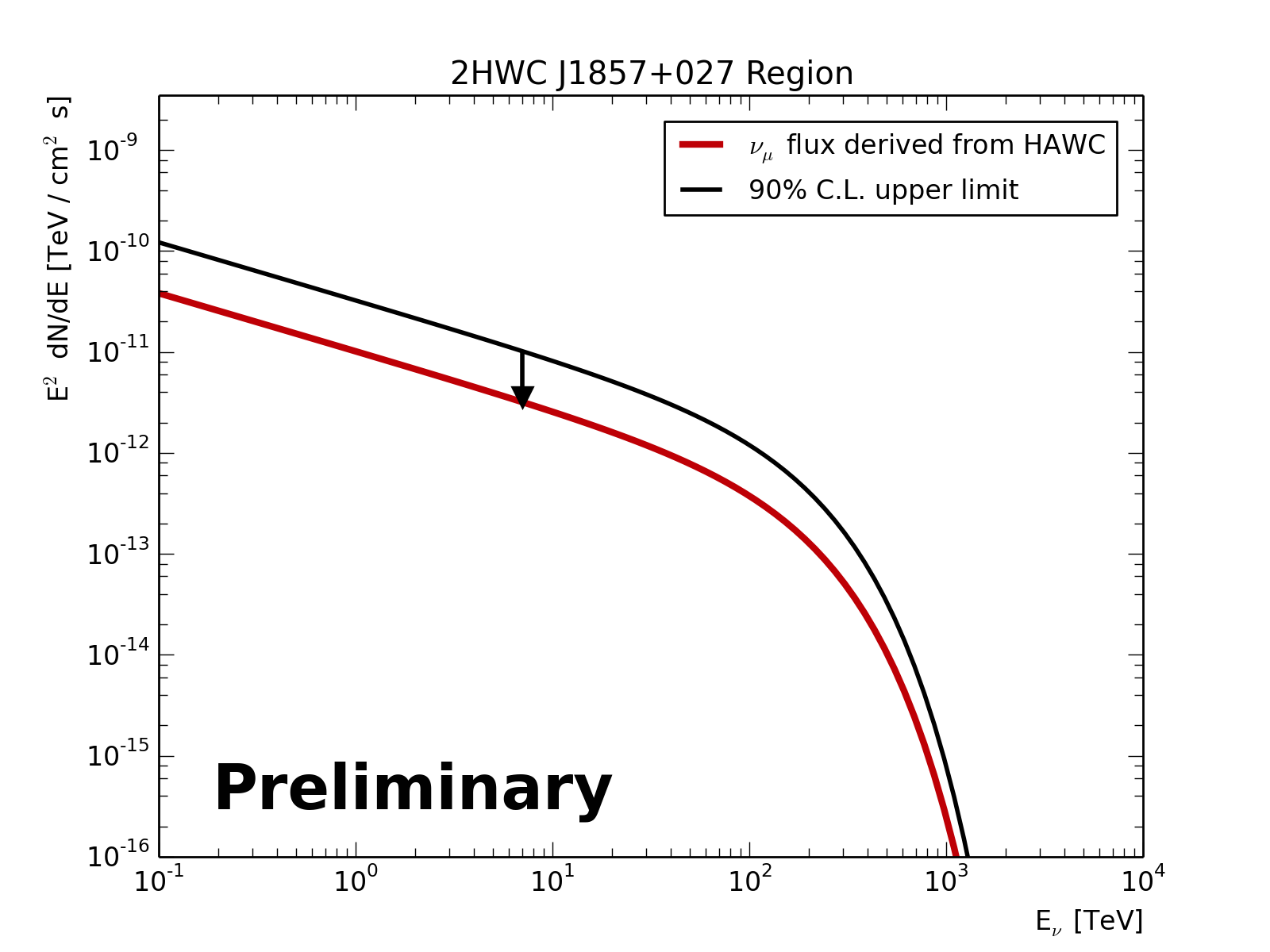}
\caption{Upper limits (90\% C.L.) on the muon neutrino flux (black) and expected high-energy muon neutrino flux (red) for the four template searches conducted. For the estimation of the neutrino flux, it is assumed that all detected high-energy gamma-rays come from a hadronic origin following \cite{Kappes:2006fg}. }
\label{fig:ul_regions}
\end{figure}

\begin{table}[t]
\begin{center}
\begin{tabular}{l|c|c|c|c}
     & Best Fit  & Sensitivity & Upper Limit (90\% C.L.) &  \\
    Search & $n_s$ & $10^{- 13}$ [TeV$^{- 1}$cm$^{- 2}$s$^{-1}$] & $10^{- 13}$ [TeV$^{- 1}$cm$^{- 2}$s$^{-1}$] & p-value \\
    \hline
     Stacking & 15.4 & 0.7 (0.6) & 1.5 (1.3) & 0.09 \\
     Northern Plane & 77.8 & 2.5 (0.4) & 5.7 (0.8) & 0.06 \\
     Cygnus Region & 0.0 & 1.0 (0.6) & 0.4 (0.2) & 0.80 \\
     J1908+063 Region & 12.0 & 0.7 (0.7) & 1.3 (1.3) & 0.14 \\
     J1857+027 Region & 36.7 & 0.8 (1.2) & 2.1 (3.2) & 0.02 \\
\end{tabular}
\caption{Results of the five analyses performed to test the correlation of high-energy neutrinos with gamma-rays observed by HAWC. The first column lists the source hypothesis being tested and the second column shows the best fit number of neutrinos from the source. The third and fourth columns show the sensitivity flux at 7 TeV and 90\% C.L. upper limit flux at 7 TeV, respectively. The final column displays the pre-trial p-value observed in each test.} 
 \label{tab:results}
\end{center}
\end{table}
In the absence of a significant excess we set upper limits on the muon neutrino flux for each of the five analyses. Fig. \ref{fig:ul_stacked_plane} and \ref{fig:ul_regions} summarize the upper limit at 90\% C.L. on $\nu_\mu+\bar{\nu}_\mu$ flux. We project the equivalent neutrino flux of HAWC's measured gamma-ray flux, assuming that all gamma-rays are hadronic. The neutrino flux is derived from the gamma-ray spectrum measured by HAWC following \cite{Kappes:2006fg}, assuming hadronuclear interaction at the sources.

The most stringent limit is set for neutrino emission from the Cygnus region. Our search found an under fluctuation of background from this region. Following IceCube convention, we have set the 90\% C.L. upper limit to the sensitivity of the test. Constructing the upper limit from the fitted values will reduce the neutrino flux from the Cygnus region by 60\%.

The upper limits on the two other regions, however, do not constrain a possible hadronic component. While the upper limit for the region surrounding 2HWC J1908+06 is marginal to the maximal hadronic emission from the source, the excess found for the 2HWC J1857+027 region yields an upper limit on the neutrino emission much larger than the corresponding gamma-ray emission.

\section{Summary \& Discussion}\label{sec:sum}

Galactic cosmic ray accelerators have long been anticipated to contribute to the flux of high-energy cosmic rays arriving at Earth. The rich environment of the Milky Way provides the targets for CRs from these sources to interact and produce high-energy neutrinos and gamma rays. Given the short distance compared to the extragalactic sources, very high energy gamma rays may survive absorption and reach the Earth. The rational for identifying Galactic sources of cosmic neutrinos relies on the multimessenger observation of both high-energy neutrinos and gamma-rays. These observations allow for a synergistic search with neutrinos and gamma rays.

The HAWC gamma-ray observatory has surveyed the Galaxy at the highest photon energies to date. Here, we used this survey to search for the Galactic origins of the high-energy cosmic neutrinos discovered in IceCube. In the absence of a significant excess from our tested hypotheses, we have determined upper limits on potential neutrino emission for each hypothesis. Therefore, obtaining the level of contribution of hadronic interactions to the high-energy emission.

 The results reported here are marginal to the expectation, and in some cases impose constraints on the hadronic flux component. The stacking and Northern plane search for neutrino emission provides a global insight to the hadronic component of the high-energy emission. On the other hand, the outcomes of examining  three specific source regions reveal important information about some of the most attractive regions in the Galaxy. The inner galaxy, the area surrounding MGRO J1908 and 2HWC J1857+027, is rich in high-energy emission but the nature of particle acceleration is not yet understood. The Cygnus region is highly complex area with relatively high level of gamma-ray emission which has been very difficult to study due to the extension of the sources. 

\newpage

\bibliographystyle{ICRC}
\bibliography{references}
\end{document}